\documentclass[12pt]{article}
\usepackage{amsmath,amssymb,graphicx,mathrsfs,slashed}
\usepackage[usenames, dvipsnames]{color} 
\usepackage{tabularx}
\usepackage[colorlinks=true]{hyperref}
\hypersetup{
	colorlinks=true,
	linkcolor=blue,
	filecolor=magenta,
	urlcolor=cyan,
	pdftitle={Sharelatex Example},
	bookmarks=true,
	citecolor=black,
}
\newcommand{\be}{\begin{equation}}
        \newcommand{\ee}{\end{equation}}
\newcommand{\bea}{\begin{eqnarray}}
        \newcommand{\eea}{\end{eqnarray}}

\def\({\left(} \def\){\right)}

\begin{document}
\title{\vspace{-1.8in}
{ Emission Channels from Perturbed\\ Quantum Black Holes }}

\author{\large Ram Brustein ${}^{(1,2)}$, Yotam Sherf ${}^{(1)}$
\\
\vspace{-.5in}  \vbox{
\begin{center}
$^{\textrm{\normalsize
(1)\ Department of Physics, Ben-Gurion University,
Beer-Sheva 84105, Israel}}$
$^{\textrm{\normalsize
(2)\ Theoretical Physics Department, CERN, 1211 Geneva 23, Switzerland}}\ \ $
\\ \small 
ramyb@bgu.ac.il,\ sherfyo@post.bgu.ac.il
\end{center}
}}
\date{}
\maketitle
\begin{abstract}
We calculate the emission of gravitational waves, gravitons, photons and neutrinos from a perturbed Schwarzschild blackhole (BH). The perturbation can be due to either  classical  or quantum sources and therefore the injected energy can be either positive or negative. The emission can be  classical in nature, as in the case of gravitational waves, or of quantum nature, for gravitons and the additional fields. We first set up the theoretical framework for calculating the emission by treating the case of a minimally coupled scalar field and then present the results for the other fields. We perform the calculations in the horizon-locking gauge in which the BH horizon is deformed, following similar calculations of tidal deformations of BH horizons. The classical emission can be interpreted as due to a partial exposure of a nonempty BH interior, while the quantum emission can be interpreted as an increased Hawking radiation flux due to the partial exposure of the BH interior.  The degree of exposure of the BH interior is proportional to the magnitude of the injected null energy.

\end{abstract}

\newpage
\renewcommand{\baselinestretch}{1.5}\normalsize

\begin{subequations}
\renewcommand{\theequation}{\theparentequation.\arabic{equation}}

\section{Introduction}

Black-Holes (BHs) are well understood in the framework of general relativity (GR) where their semiclassical properties in the exterior region have been successfully described using the formalism of quantum fields in curved spacetime. In contrast, their quantum nature, in particular, quantum effects in the vicinity of the horizon, has not yet been fully understood and is inconsistent with the classical GR description. For a review, see, for example, \cite{Harlow:2014yka,Polchinski:2016hrw}.

Hawking \cite{Hawking:1974sw} considered BHs in equilibrium and demonstrated that  BHs emit particles as if they were a thermal body with a temperature $T_H$ that is inversely proportional to their mass $M$.  The established method of studying quantum BHs is the semiclassical approach, in which the background  gravitational field is fixed and matter fields are quantized about this fixed background. The quantization of matter fields in Schwarzschild spacetime and particularly the vacuum expectation value (VEV) of the stress energy momentum (SEM) tensor was calculated in \cite{Dowker:1975tf,Hawking:1976ja,Brown:1976wc,Christensen:1976vb,Christensen:1978yd,Candelas:1980zt,Page:1982fm}. The inherent divergence of the VEV in curved space was resolved  by applying an appropriate regularization technique. These results emphasize that creation of particles in curved space is a vacuum phenomenon: virtual particles gain sufficient energy from the background so as to become real.

Here, our main objective is to examine the properties of BHs by examining their emission when they are away from equilibrium. This is achieved by considering external perturbations to the BH. Specifically, we calculate modifications to Hawking radiation and to gravitational wave (GW) emissions. Of particular interest are perturbations that deform the BH horizon, thus allowing it to be deformed inwards. In general, these require absorption of some negative null energy by the BH and therefore are likely to result from quantum processes. However, some regions of the BH horizon can be deformed inwards if the total injected nukk energy is small. We show that, unlike the gravitational perturbations in classical GR, where radiation can only be emitted from a Schwarzschild BH in the form of gravitational waves, in the quantum case, due to coupling of the background metric to the various matter fields and their nonzero VEV of the SEM tensor, gravitational perturbations produce additional particle species. However, the rate of such particle production is small.

The paper is organized as follows, in the first part we establish the theoretical framework for evaluating the quantum emission from perturbed BHs. We derive an explicit expression for the radiated power by using the path integral approach and show that it coincides with the Euclidean partition function approach. In the second part we outline the classical setup that describes the deformed horizon geometry of BH in the presence of external perturbation and determine their corresponding metric perturbation. In the third part we apply the theoretical framework and calculate the quantum emission for various particle species. The classical emission is calculated by following the relevant discussions on tidal deformations in the literature. Next, we compare the results and discuss their relative magnitude in the context of horizon deformations. We find that the relative magnitude is controlled by the BH entropy and by the time scale of the emission. The factor involving the time scale can, in some cases provide a significant enhancement in the emitted flux. Then, we interpret the increased flux in terms of an external observer and a nonempty BH. In the final part, we summarize our result and discuss their significance. In an appendix, for completeness,  we explain in detail the relationship between our discussion of tidal deformations and the existing discussions in the literature.

\end{subequations}

\begin{subequations}
\renewcommand{\theequation}{\theparentequation.\arabic{equation}}

\section{Emission from perturbed blackholes}\label{subsection1.1}

In this section we present the formalism for describing how external perturbations modify the vacuum state of quantum fields outside the BH. The modified state is time dependent and so, according to standard arguments, particle production occurs.

First, we consider the simple case of a minimally coupled, massless scalar field, whose equations of motion are given by $g_{ab}\nabla^a \nabla^b \phi=0$, where $g_{ab}$ is the unperturbed background Schwarzschild metric. In the absence of perturbations, when the field is in its vacuum state, the vacuum persistence amplitude  $\langle\text{out},0|0, \text{in} \rangle=Z[0]$, is given in terms of an effective action,
\begin{equation}
\begin{split}
Z[0] ~=~\int \mathcal{D}[\phi] ~e^{iS_M[\phi]}~,
\end{split}
\end{equation}
where the scalar fields action is given by
\begin{gather}
S_M~=~\frac{1}{2}\int d^4x \sqrt{-g} g_{ab}\nabla^a \phi \nabla ^b \phi~.
\end{gather}
 In general, in curved spacetime  the “in” and “out” vacuum states are different, this is also the case for time-independent backgrounds, such as Schwarzs-\\child spacetime. In a curved background the annihilation operators for the past ``in" vacuum, $\hat{a}_j$ are defined as $\hat{a}_j|0, \text{in} \rangle=0$ while the corresponding operators of the future ``out" region $\hat{b}_j$, are defined as $\hat{b}_j|0, \text{out} \rangle=0$. Generically, the operators $\hat{a}_j$ are different than $\hat{b}_j$ and the two sets are related by a Bogolubov transformation (for more details, see for example \cite{D,B,DeWitt:1975ys} and in the context of this manuscript, see explicitly in \cite{Sherf:2019arn}). In particular, also in Schwarzschild spacetime  $\hat{b}_j|0, \text{in} \rangle\neq 0$ and $\hat{a}_j|0, \text{out} \rangle\neq 0$ which implies that $\langle	\text{out},0|0, \text{in} \rangle \neq 0$ and therefore that particles are being produced.

To investigate the influence of perturbations on the vacuum state, we consider a finite-duration small perturbation to the background Schwarzschild metric. The perturbation is switched on at some early time $t_i$ and switched off at some later time $t_f$. So, the metric changes from  $g_{ab}$ to $g_{ab}+h_{ab}$. The scalar field action variation is related to stress-energy-momentum tensor,
\begin{gather}
{\delta S_M}~=~\dfrac{\sqrt{-g}}{2} h_{ab}T^{ab},
\label{SM}
\end{gather}
where $T^{ab}$ is the SEM tensor of the minimally coupled field given by
\begin{gather}
T^{ab}~=~\dfrac{1}{2}\nabla^a\phi\nabla^b\phi-\dfrac{1}{4}g^{ab}\nabla^c\phi\nabla_c\phi.
\label{1.13}
\end{gather}
The perturbation makes the generating functional time dependent. The time-dependent generating functional $Z[h]$ depends on the metric perturbation $h_{ab}$. The functional $Z[h]$ can be regarded as the generating functional in the presence of a driving source term $h_{ab}$ and is written as follows,

\begin{equation}
\begin{split}
Z[h]= \int \mathcal{D}[\phi] ~e^{iS_M[\phi]}e^{\frac{i}{2}\int d^4x {\sqrt{-g}}h_{ab}T^{ab}}
\label{VPAT}
\end{split}
\end{equation}
Following the semiclassical approach we define the effective action $\langle	\text{out},0|0, \text{in} \rangle=e^{i\mathcal{W}}$ and apply the relation
\begin{equation}
\begin{split}
{\int \mathcal{D}[\phi] ~e^{iS_M[\phi]}e^{\frac{i}{2}\int d^4x {\sqrt{-g}}h_{ab}T^{ab}}}~=&~{e^{i\mathcal{W}}}~	\langle e^{\frac{i}{2}\int d^4x {\sqrt{-g}}h_{ab}T^{ab}}\rangle ~.
\end{split}
\end{equation}
We then approximate
\begin{equation}
\begin{split}
e^{i\mathcal{W}}~\langle e^{\frac{i}{2}\int d^4x {\sqrt{-g}}h_{ab}T^{ab}}\rangle
&\approx~~e^{i\mathcal{W}}~e^{\frac{i}{2}\int d^4x {\sqrt{-g}}h_{ab}\langle T^{ab}\rangle}~.
\end{split}
\label{zhlin}
\end{equation}

 A naive calculation of $\langle T^{ab}\rangle$ is divergent.
Therefore, we have to consider the VEV of the renormalized stress-energy-momentum (RSEM) tensor, denoted by $\langle T^{ab}\rangle_{\text{ren}}$. There are many renormalization techniques that handle the infinities of $\langle T^{ab}\rangle$ in curved space.
While in flat backgrounds divergences of the SEM tensor VEV are easily identified and eliminated by the standard normal ordering techniques, in curved backgrounds this method is inapplicable and rather intricate regularization techniques are introduced. The most frequently used among them are the zeta function regularization \cite{Dowker:1975tf}, dimensional regularization \cite{Hawking:1976ja,Brown:1976wc} and the  covariant geodesic point separation \cite{Christensen:1976vb,Christensen:1978yd,Candelas:1980zt,Page:1982fm}, which we elaborate on bellow.

In general, in curved background the full partition function is given by $S=S_g+S_m$ where $S_g$ is the  gravitational Einstein-Hilbert action. Then, as shown above, the external source  induces a gravity-matter coupling of the form $h_{ab}\langle T^{ab}\rangle$. Following the separation method \cite{Christensen:1976vb,Christensen:1978yd,Candelas:1980zt,Page:1982fm}, the divergences in the SEM tensor are isolated such that $\langle T^{ab} \rangle =\langle T^{ab}\rangle_{\text{ren}}+\langle T^{ab}\rangle_{\text{div}}$, where the divergent terms in $\langle T^{ab}\rangle_{\text{div}}$ turn out to be purely geometrical and can be absorbed into the gravitational action. The finite $\langle T^{ab}\rangle_{\text{ren}}$  determines the emission of radiation from the BH.

The final result for $Z[h]$ is obtained by replacing $\langle T^{ab} \rangle$ by  $\langle T^{ab}\rangle_{\text{ren}}$ in Eq.~(\ref{zhlin}),
\begin{equation}
Z[h]~
\approx~~e^{i\mathcal{W}}~e^{\frac{i}{2}\int d^4x {\sqrt{-g}}h_{ab}\langle T^{ab}\rangle_{\text{ren}}}~.
\end{equation}
It is also instructive to examine the first order expansion of $Z[h]$,
\begin{gather}
Z[h]~\approx ~e^{i\mathcal{W}}\left(1+\frac{i}{2}\int_0^t dt' \int d^3x {\sqrt{-g}}h_{ab}\langle T^{ab}\rangle_{\text{ren}}\right) ,
\label{ZH}
\end{gather}
where now it is clear that the first term is the Hawking term that gives rise to the thermal flux, while the second term is a modification of the Hawking radiation induced by the external perturbation.

It is convenient to work in the interaction picture, assigning the explicit time-dependence of the ``in" state to the corresponding operator,
\begin{equation}
\begin{split}
|0,\text{in} \rangle_t~&=~e^{\frac{i}{2}\int_0^t dt'\left(\int d^3x {\sqrt{-g}}h_{ab}T^{ab}\right)}|0, \text{in} \rangle~
\label{1.53}
\end{split}
\end{equation}

 In this form, we identify the time-evolution operator $U(t,0)|0,\text{in}\rangle=|0,\text{in}\rangle_t$, and the energy difference $\Delta E$ between the states $|0, \text{in}\rangle $ and $|0, \text{in} \rangle_t$, $ |0,\text{in}\rangle_t=e^{-i\int_{0}^{t}\langle \Delta E \rangle dt'}|0,\text{in}\rangle$
 \begin{gather}
  \langle \Delta E \rangle~=~-\frac{1}{2}\int d^3x {\sqrt{-g}}h_{ab}\langle T^{ab}\rangle~.
 \label{ddd1}
 \end{gather}

Equation~(\ref{ddd1}) can also derived by using the Euclidean partition function.
The Euclidean rotation is valid as long as the relative corrections to the background Schwarzschild geometry are small. Then, the corrections to the periodicity of the Euclidean time $\beta$ are small\footnote{The corrections to $\beta$  due to the external perturbation result in an additional second order correction in Eq.~(\ref{ddd1}).}. Furthermore, the perturbations that we consider do not induce any additional singularities.

The Euclidean energy is given by  $\langle E \rangle = -\frac{\partial \text{ln}Z}{\partial \beta}$. The energy difference between the perturbed and the unperturbed states $ \langle  \Delta E \rangle=\langle E(t) \rangle-\langle E \rangle$, where $\langle E(t) \rangle, \langle E \rangle$ are derived from their associated partition functions $Z[h], Z[0]$ in accordance. Then, $\langle  \Delta E \rangle = -{\frac{1}{2}\int d^3x {\sqrt{-g}}h_{ab}\langle T^{ab}\rangle_{\text{ren}}}$ which agrees with Eq.~(\ref{ddd1}).

The excess energy of the time-dependent vacuum state with respect to the stationary initial vacuum state $\langle  \Delta E \rangle$, which is supplied by the external gravitational perturbation is  the total energy gained by the vacuum state of the matter fields. Part of this energy, which is denoted by $\Delta E_R$, is emitted to infinity in different forms of radiation and part remains trapped and contributes to the eventual increase in the mass of the BH.  The distribution of the radiated energy depends on the detailed nature of the perturbation and the type of the excited fields. We are particularly interested in the total power radiated by the $l=2$ deformation of the horizon surface (labeled by $R_{2}$) or equivalently the BH surface luminosity. This power is obtained by projecting the flux components of the SEM tensor on the BH surface, which is defined by embedding the BH's outer surface in a constant time surface $\Sigma_t$. The flux of the SEM tensor is given by the projection $T^{t\rho}n_{\rho}=T^{tr}$, where $n_\rho=(0,1,0,0)$ is the normal to $\Sigma_t$. Then, from Eq.~(2.9) the radiated energy to infinity as a result of deformations in the BH outer surface is given by $\Delta E_R = -{\frac{1}{2}\int d^3x {\sqrt{-g}}h_{tr}\langle T^{tr}\rangle_{\text{ren}}}$ and the additional power emanating from the deformed horizon (the BH additional luminosity) takes the form
 \begin{gather}
  \Delta L~=~-\frac{1}{2}\int_{\Sigma_t} d\Omega r^2h_{tr}\langle T^{tr}\rangle_{\text{ren}},
 \label{1.71}
 \end{gather}
 where $\Delta L$ is the luminosity difference between the stationary and the time-dependent state.
 Moreover, by identifying the Hawking luminosity as $L_H=\int_{R=2M} d \Omega r^2 \langle T^{tr}\rangle_{\text{ren}}$,  Eq.(\ref{1.71}) can be written as $\Delta L \sim  h_{tr}{(R_{lm})}L_H$,  where $h_{tr}{(R_{lm})}=h_{tr}(r)\big|_{r\rightarrow R_{2m}}$ is the value that the metric perturbation takes about the deformed horizon.

 In the following, we outline the perturbative treatment and specify the metric perturbation that describes geometric deviations of the BH horizon away from, but near, equilibrium.	

\end{subequations}
 \begin{subequations}
 	\renewcommand{\theequation}{\theparentequation.\arabic{equation}}

\end{subequations}

\begin{subequations}
\renewcommand{\theequation}{\theparentequation.\arabic{equation}}
\section{Classical setup}

As previously mentioned, our interest is in the emitted radiation from BHs that are out of equilibrium. BHs out of equilibrium were discussed in the context of studying tidally deformed BHs. There, the perturbations are described in the horizon-locking gauge \cite{Poisson:2004cw,Poisson:2009qj,Vega:2011ue,Poisson:2018qqd,OSullivan:2014ywd}. We have adopted this framework for our purposes. We first outline some of the main ideas and then discuss the detailed calculations and results.

\subsection{Background}\label{BD}
Originally, the dynamics of BHs undergoing tidal deformations was described by Thorne and Hartle \cite{Thorne:1980ru} and later elaborated by  Alvi \cite{Alvi:1999cw}, Hughes \cite{OSullivan:2014ywd} and Poisson \cite{Poisson:2004cw,Poisson:2009qj,Poisson:2018qqd,Taylor:2008xy} who describes in great detail the geometry of a deformed BH horizon as a result of tidal gravitational perturbation.
The idea is that the spcetime of a nonrotating BH is deformed by a weak tidal interaction produced by an external moving object. This environment is characterized by the radius of curvature $\mathcal{R}$, which can be considered as the region in space where the BH gravitational field interacts with the object's tidal field. To guarantee a weak gravitational interaction, it is assumed that the Schwarzschild BH with mass $M$ and radius $R_S=2GM$ satisfies $R_S\ll\mathcal{R}$. For example, consider a binary system that consists of a BH and an external  object with mass $M'$, the relative distance between them is $b$. Then, the radius of curvature is given by $\mathcal{R}^2\sim b^3/(M+M')$, which is of order of the BH's angular velocity, which also determines the typical interaction scale. The important aspect is  that the tidal field induces a modification to the BH gravitational field in a region $r\ll\mathcal{R}$. So, perturbation about the background geometry are expanded in powers of the dimensionless parameter $r/\mathcal{R}\ll1$.
For a more detailed description we refer the readers to Appendix \ref{AX}, and to Ref.~ \cite{Vega:2011ue}. In the following section, in analogy with Poisson's results, we present a general framework for constructing the geometry of deformed horizons.

\subsection{Deformed horizon geometry}\label{DHG}
As previously mentioned, we are interested in  perturbations that describe a horizon deformation of BHs near their equilibrium state. Astrophysical BHs relaxing to equilibrium can originate from a various astronomical events such as a BH that is immersed in an external gravitational field induced by an external source or by inspiralling compact binaries \cite{Poisson:2009qj}-\cite{Blanchet:2013haa}, or by a binary postmerger event in its ringdown stage \cite{Buonanno:2006ui,Price:1994pm}. Here the specific details of these events are not important, instead we are interested in the general description of the horizon deformation.

The idea is that an external GR observer can describe the near-horizon geometry of a deformed BH relaxing to equilibrium in the horizon-locking coordinate system. In this gauge, the horizon position is ``locked" at $r=R_S$, such that $h(R_S)=0$ up to some higher-order correction in the perturbation strength. Then the perturbed geometry is interpreted in terms of a perturbation in the associated Ricci curvature. Alternatively, the deviation in the scalar curvature can be converted into a deviation of the BH outer surface. An external observer can interpret the perturbation as the geometrical deviation of the BH outer surface horizon from its unperturbed location at $r=R_S$ \cite{Vega:2011ue,OSullivan:2014ywd}. This is depicted in Fig.~\ref{F} for some specific perturbations. In general, outward and inward deviations of the BH surface with respect to its unperturbed horizon $R_S$ can be  identified as the injection of some average positive and negative null energy, respectively.

To determine the deformed horizon geometry, the BH outer surface position is parametrized about a three-dimensional Euclidean space as measured by a remote flat space observer. The dimensionless horizon displacement parameter of the deformed surface with respect to its unperturbed position is denoted by $D(v)_{lm}$. Here $v$ is the ingoing Eddington-Finkelstein (EF) coordinate , and $l,m$ are the deformation angular modes. Then $D(v)_{lm}\equiv D(v)Y_{lm}$ and specifically for the $l=2$ mode we define
\begin{gather}
	D_2~=~D \sum_m Y_{2m}.
	\label{DQ2}
\end{gather}
The BH outer surface position for an arbitrary $l,m$ modes is given in this case by,
\begin{gather}
R(v)_{lm}~=~R_S\large(1+D(v)_{lm}\large)~,
\label{Rlm}
\end{gather}
in accordance with Eq.~({\ref{DQ2}}), the surface displacement for the $l=2$ mode is $R_2=R_S(1+D_2)$.
This is illustrated in Fig.~\ref{F} for some $l,m$ modes.
As stated, this parametrization defines the BH surface in a three-dimensional Euclidean space. The surface curvature associated with the above parametric equation is given by
\begin{gather}
\mathcal{R}~=~\dfrac{1}{2M^2}\left(1+4D Y_{lm}\right)~,
\label{RR}
\end{gather}
where the $Y_{lm}$ are defined as a real functions (see Appendix \ref{AX}, Eq.~(\ref{SPT})). To proceed, we identify the surface curvature with the Ricci curvature of the deformed horizon geometry. Then, since the only information that we have is about the surface curvature, we have some gauge freedom in determining uniquely the near-horizon geometry. In \cite{Poisson:2004cw,Poisson:2009qj} it is shown that there exists a unique choice of coordinates that satisfies the geometrical properties mentioned above, the horizon-locking coordinate system.

  \begin{figure*}[h!]
	\hspace{-2.12cm}
	\includegraphics[scale=.55]{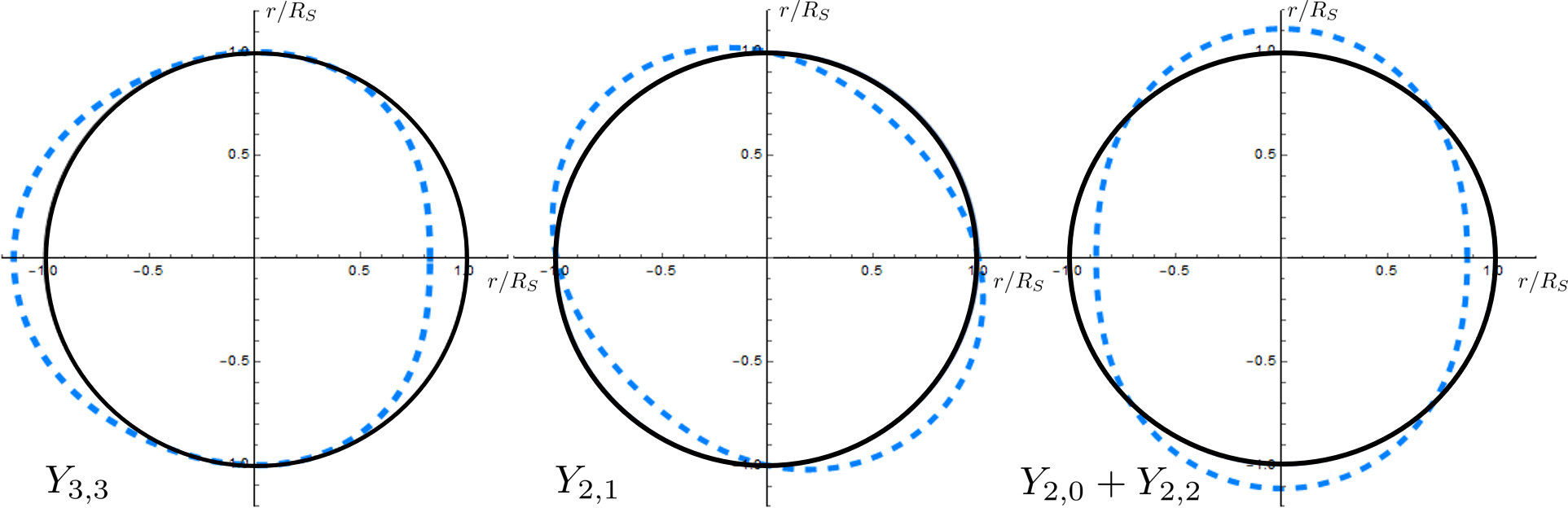}
	
	\caption[{{\small The goemetric deviation of the BH outer surface for the differnt mode functions}}]{{\small \vspace{-0.2cm}The geometric deviation of the BH outer surface for the different mode \vspace{-0.2cm}functions: The solid (black) line is the unperturbed BH or equivalently the perturbed BH in the horizon-locking coordinates. The dashed (blue) is the position\vspace{-0.2cm} of the deformed BH horizon in Schwarzschild coordinates in the presence of\vspace{-0.2cm} perturbations. }}
	\label{F}
\end{figure*}

The  horizon-locking  metric perturbation is defined about the Schwarzschild background in the outgoing EF coordinates whose line element is given by
\begin{gather}
ds^2~=~-f(r)du^2-2dudr+r^2d\Omega^2~,
\label{ED}
\end{gather}
where $f(r)=1-2M/r$. The metric perturbation is given as an expansion in the dimensionless horizon displacement parameter $D<1$. Here we give only the first nonvanishing term which is of order $D$. This term, as explained in  Appendix \ref{AX} and also in \cite{Poisson:2004cw}, is the contribution of the quadrupole moment $l=2$. The higher-order terms are the octupole moment $l=3$ of order $\sim D^{3/2}$ and the $l=4$ hexadecapole moment $\sim D^{2}$. Furthermore, since we are interested in calculating the deformed BH luminosity, given by Eq. (\ref{1.71}), the only relevant metric component is given in Schwarzschild coordinates by $h_{tr}$ or in EF coordinates by $h_{uu}$. The other metric components do not contribute to the radiated flux, rather they do participates in the time-evolution of the initial vacuum state given in Eq.~(\ref{1.53}). We listed them explicitly in Appendix \ref{AX}.  Then, in the vicinity\footnote{An exact definition of the ``vicinity region" appears in Appendix \ref{AX}.} of the BH horizon, the metric perturbation for the $l=2$ mode Eq.~(\ref{DQ2}) takes the form
\begin{equation}
h_{uu}~=~4\left(1-\dfrac{r}{2M}\right)^2D_{2}~.
\label{HU}
\end{equation}

Now, when the theoretical framework for the evaluation of the classical and quantum emission is established,
we are able to perform the desired calculations.

							\section{Results}

Here, we begin with the results that describe the classical gravitational wave emission from a perturbed Schwarzschild BH. We then proceed to calculate the quantum emission of fields with different spin: minimally coupled conformal scalar fields, electromagnetic fields, neutrino fields and graviton fields. This emphasizes that the framework for the classical and quantum emission is similar while demonstrating the fact that unlike the classical emission, in the case of quantum emission, all matter fields are produced.

					\subsection{Quadrupole emission}\label{QE}

			The deformation of the Schwarzschild BH horizon is given in Eq.~(\ref{Rlm}) and its associated background perturbation in Eq.~(\ref{HU}). This deformation leads to a nonvanishing time-dependent quadrupole moment as we now explain. The time-dependent geometric deviation results in a time-dependent excess energy density $\rho(t,r)$. This leads to a time-varying quadrupole moment $Q_{ab}\sim \int d ^3 x' \rho(t,x')x_ax_b$, which in turn sources a gravitational-wave emission. It is possible to express the stress tensor of the gravitational energy and its radiation power in the standard form $\Delta L _{GW}\sim \langle \dddot{Q}_{ab}\dddot{Q}^{ab} \rangle$. This can be estimated directly by noting that $Q\sim M^3$. As explained in Appendix \ref{AX}, we assume that the induced gravitational field is slowly varying, then its rate of change can be approximated by  $d/dt\sim\Delta t^{-1}\sim  D^{1/2}/M$, so $\dddot{Q}^2\sim D^3$ and $\Delta L_{GW}\sim D^3$. Detailed calculations of the classical emission from a BH undergoing tidal deformations are found in \cite{Poisson:2004cw,Thorne:1980ru}.
			The final result for the emitted power of gravitational energy, or alternatively the rate in which the BH losses its mass, is expressed in terms of the dimensionless displacement quantity $D$, and takes the  from
			\begin{equation}
			\begin{split}
			\Delta L _{GW}~\approx~\dfrac{c^5}{G} \left(\dfrac{T_S}{\Delta t}\right)^2D^2,
			\end{split}
			\label{PC}
			\end{equation}
			where $\Delta t$ is the characteristic time interval of the perturbation is denoted by $\Delta t$ and $T_S$ is the Schwarzschild time $T_S=R_S/c$. Then, if the time interval scales as  $\Delta t \sim M/D ^{1/2}$, then $T_S/\Delta t\sim D^{1/2}$ and the power becomes $\Delta L_{GW}\sim D^3$.
			
			The total energy loss in a characteristics time interval $\Delta t$ is  $\Delta E_{GW}=\Delta L _{GW}\Delta t$
			\begin{gather}
			\Delta E_{GW} \approx \dfrac{c^5}{G} \dfrac{T_S^2}{\Delta t}D^2
			\end{gather}

In the models that were considered in \cite{collision} the perturbations are induced by exciting exotic matter in the interior of the BH. In this case, the scaling of the various quantities is different than the one discussed above. The scaling of the frequency is $\omega \sim D$ and that of the lifetime is $\tau \sim 1/D^2$, so $\Delta L_{GW} \sim D^4$ and $\Delta E_{GW}\sim D^2 M_{BH}$. In this case, the emitted energy is a finite fraction of the BH energy and therefore is detectable, in principle, by gravitational-wave experiments. If the BHs are rotating, the amount of emitted energy can further increase beyond this estimate.

			\subsection{Increased quantum flux}\label{iqf}
	
			To begin, we specify the quantities in the expression for the radiated energy flux, Eq.~(\ref{1.71}),
			recalling that we are interested in the emitted power from the deformed horizon.
			
		The flux components of the RSEM tensor are evaluated in EF coordinates about the Unruh vacuum state, which describes an outward thermal flux at infinity. Then, the flux term is given by $T_{uu}-T_{vv}=2T_{tr}$ so the Hawking luminosity for a field of a given spin is defined as $L_H^s~=~4\pi r^2\langle T_{tr}\rangle_{\text{ren}}^s$. The explicit expression for the RSEM tensor of a stationary Schwarzschild BH is given by \cite{Davies:1976ei},
		\begin{gather}
		L_H^s~=~\bar{\alpha} ^s\dfrac{\pi}{12}T_H^2~,
			\label{ttr}
		\end{gather}
			 with $T_H=1/(8\pi M)$ being the Hawking temperature \footnote{To restore units, the luminosity needs to be multiplied by a factor of $\frac{k_B^2}{\hbar}$.}. The numerical factor $\bar{\alpha}^s$ depends on the spin of the excited fields and is determined by the transmission coefficient through potential barrier \cite{Page:1976df}. Equation~(\ref{ttr}) describes a constant thermal flux at the Hawking temperature $T_H$. The metric perturbation in Eq.~(\ref{HU}) in Schwarzschild coordinates is given by $h_{tr}=f(r)^{-1}h_{uu}$,
			\begin{gather}
				h_{tr}~=~-D_{2}\dfrac{r^2}{M^2}\left(1-\dfrac{2M}{r}\right)~.
				\label{htr}
			\end{gather}
			Finally, we calculate the additional radiation power emanating from the deformed horizon at ${r= R_{2}}$, where according to Eq.~({\ref{Rlm}}) the BH outer surface is located at $R_{2}=R_S(1+D_{2})$. Recall that $D_{2}=D\sum_m Y_{2m}$, then the BH surface luminosity for the dominant quadrupole $l=2$ perturbation is given by Eq.~(\ref{1.71})
			\begin{equation}
			\begin{split}
			\Delta L^s~&=-\frac{1}{2}\int_{\Sigma_t} d\Omega r^2h_{tr}\langle T^{tr}\rangle^s_{\text{ren}}\Big|_{r\rightarrow R_{2}}~\\~&=-\frac{1}{2}\int_{\Sigma_t} d\Omega r^2\left[D_{2}\dfrac{r^2}{M^2}\left(1-\dfrac{2M}{r}\right) \left(\bar{\alpha}^s\dfrac{T_H^2}{48r^2}\right)\right]\bigg|_{r\rightarrow R_{2}}~.
			\end{split}
			\end{equation}
			By substituting $M=1/8\pi T_H, R_S=2M$ and ${r\rightarrow R_{2}}$ we obtain the following expression,
			\begin{equation}
			\begin{split}
			\Delta L^s~=&~\dfrac{2\pi^2}{3}{\bar{\alpha}^s}{T_H^4}\int_{\Sigma_t}D_2 d\Omega\left(r^2-rR_{S}\right)\Big|_{r\rightarrow R_{2}}\\
			\approx&~
			\dfrac{{\alpha}^s}{24}D^2 T_H^2\int_{\Sigma_t} d\Omega \sum_{m}(Y_{2m})^2+O(D^3)~.
			\label{DLS}
			\end{split}
			\end{equation}
			In passing from the first to the second line we substituted $D_{2}=D\sum_m Y_{2m}$.
			Finally, we substitute the corresponding $l=2$ spherical harmonics that are defined in the Appendix Eq.~(A.10) and perform the azimuthal integral $\int d\Omega \sum_m (Y_{2m})^2$. This eventually yields
			\begin{gather}
			\Delta L^s~\approx~\dfrac{14 \pi}{45}\bar{\alpha}^s(D T_H)^2~.
			\end{gather}
			First, it is clear that the surface luminosity increases with the horizon displacement parameter $D$, which means that, geometrically, a larger deformations of the BH horizon $\Delta R_S \sim DR_S$ leads to a higher luminosity.
			In addition, the flux is positive, independently of the sign of $D$ and independent on whether the BH outer surface protrudes outward or inward.

			   Equation~(\ref{DLS}) can be further simplified by denoting $L_H^s=4\pi r^2\langle T^{tr}\rangle^s_{\text{ren}}$, so  takes the form\footnote{We neglect the corrections of the spin coefficient $\bar{\alpha}^s$ since they are higher order in $D$.}
			\begin{gather}
				\Delta L^s~=~\dfrac{56}{15}D^2  L_H^s~.
				\label{LL1}
			\end{gather}
				For completeness, we list the final results, including the average spin coefficient factors \cite{Page:1976df}. Applying the definition of the spin-independent Hawking luminosity $L_H= \pi T_H^2/12 $, the excess emitted power for massless scalar fields is $\Delta L^0=0.66\times D^2 L_H$, for all species of spin half massless neutrinos (if they exist) $\Delta L^{1/2}=1.47\times D^2 L_H$, and for photons and gravitons $\Delta L^{1}=0.3\times D^2 L_H$, $\Delta L^{2}=0.03\times D^2 L_H$, respectively.
				
				This emphasizes that the relaxation of BH to equilibrium is accompanied by an emission from two different origins. The first is of quantum origin $\Delta L^s$, which is sourced by  the fluctuation of the SEM tensor. This emission is distributed between the different channels. The strength of the emission in a particular channel $s$, is determined by $\alpha^s$.
				The second emission channel, is of classical origin Sec.~\ref{QE} and is labeled by $\Delta L_{GW}$. To demonstrate how energy is radiated through the classical GW channel we derive the classical analog of Eq.~(\ref{1.71}). At first, we consider the gravitational action \begin{gather}
				S_g~=~\frac{1}{2}\int d^4x \sqrt{-g} g_{ab}\mathcal{R}^{ab}~.
				\end{gather}
				where $\mathcal{R}^{ab}$ is the Ricci tensor. Then, we derive the SEM tensor of the emitted radiation in the far-field region and expand the gravitational action in the GW perturbation $h_{ab}$. The zeroth and first order expansion of the Ricci tensor vanishes in the Schwarzschild vacuum whereas the second order expansion is nonvanishing and yields (after lengthy derivation) the null component of the emitted GWs SEM tensor
				\begin{gather}
				\delta \mathcal{R}_{uu}^{(2)}~=~	t_{uu}~=~-\dfrac{1}{8\pi r^2}\langle \dddot{Q}_{ij}\dddot{Q}^{ij} \rangle
				\end{gather}
				and the GW power is given by
				\begin{equation}
				\begin{split}
				\Delta L_{GW}~&=~-\dfrac{1}{2}\int_{\Sigma_t}d\Omega t_{uu}r^2\\
				&=~-\dfrac{1}{5}\langle \dddot{Q}_{ij}\dddot{Q}^{ij} \rangle
				\end{split}
				\end{equation}
				This demonstrates that the classical GW emission originated in the nonzero quadrupole moment that is generated by  deformations of the BH.
				
					To proceed, it is possible to estimate the modified Hawking temperature by using Eq.~(\ref{ttr}) and assuming that the perturbed BH emits an approximately thermal spectrum. Then the emission is approximately that of a black body.
				The modified luminosity $\widetilde{L}=L_H+\Delta L$, can be evaluated using Eq.~(\ref{ttr}),  $\widetilde{L}=\frac{\pi}{12}\widetilde{T}^2$, where $\widetilde{T}$ is the modified Hawking temperature of the BH surface. From Eq.~(\ref{DLS})
				\begin{gather}
				\dfrac{\pi}{12}\widetilde{T}^2~=~	\dfrac{\pi}{12}\widetilde{T}_H^2+\dfrac{14 \pi}{45}(D T_H)^2~,
				\end{gather}
				so $\widetilde{T}$ is given by
				\begin{gather}
				\widetilde{T}~=~T_H\sqrt{1+\dfrac{56}{15}D^2}~.\\
				\dfrac{\Delta T}{T_H}\approx \dfrac{28}{15}D^2~,
				\label{TM}
				\end{gather}
				where $\Delta T=\widetilde{T}-T_H$.
				The modified BH temperature is seen no to depend on the sign of $D$, which implies that inward or outward surface deviations Eq.~(\ref{Rlm}) resulted in an increase of the BH temperature.
				
			To get more insight about the results, it is instructive to compare between the the classical and the quantum emissions, in Eq.~(\ref{PC}) and Eq.~(\ref{LL1}), respectively,
			\begin{equation}
			\begin{split}
				\dfrac{\Delta L}{\Delta L _{GW}}~&\approx~ \dfrac{k_B^2 G}{\hbar c^5} \left(\dfrac{\Delta t}{T_S}\right)^2{T_H}^2\\
				&=~ \dfrac{\hbar c}{G M^2 }\left(\dfrac{\Delta t}{8\pi M }\right)^2~=~\dfrac{1}{4\pi} \left(\dfrac{\Delta t}{T_S}\right)^2\dfrac{1 }{ S_{BH}}.
				\label{ytr}
			\end{split}\end{equation}
		The appearance of a quantum suppression factor  $1/S_{BH}$ can be explained as follows.
				Classically an unperturbed BH has energy mass of $E=M$, also, the energy scale that is associated with its quantum properties is the BH temperature $T_H=1/R_S$. Then, in analogy with the comparison of the classical to quantum emission Eq.~(\ref{ytr}), their ratio for an unperturbed BH yields
				\begin{gather}
					\dfrac{T_H}{M}=\dfrac{1}{R_SM_{BH}}=\dfrac{1}{S_{BH}}.
					\label{up}
				\end{gather}				
Therefore, we should expect a suppression factor of order $1/S_{BH}$ to appear in Eq.~(\ref{ytr}). However, note the additional factor $(\Delta t/ T_S)^2$ which, for large $\Delta t\gg T_S$, may significantly increase the relative magnitude in  Eq.~(\ref{ytr}). Here, $\Delta t$ is left unspecified and depends on the details of the underlying mechanism that drives the BH horizon deformations. For example, we can consider the setup of horizon deformations that are induced by an external remote moving object, and are given in Section (\ref{BD}) and also in Appendix \ref{AX}. The characteristic time scale is $\Delta t \sim T_S/D^{1/2}$, then the ratio exhibits $\Delta L/\Delta L _{GW}\sim 1/(D S_{BH})\sim l_P^2/(DR_S^2)$. To interpret that we recall that the radial horizon displacement is given by $\Delta R_S=D R_S$, then for small deviations with $D\ll1$ we argue that in BH relaxing to equilibrium, the ratio Eq.~(\ref{ytr}) is considerably larger in comparison with the ratio of an unperturbed BH Eq.~(\ref{up}). Furthermore for the sub-Planckian regime where $D R_S^2\sim l_p^2$, which means that $\Delta R_S\sim l^2_p/R_S\sim R_S/S_{BH}$, we conclude that for deviation of this order the quantum emission is nonnegligible in comparison to the classical one.

			\section{Summary and Discussion}
			
In this paper we calculated the emission  from a perturbed Schwarzschild BH. The perturbations could be  of  classical or quantum origin . We established a theoretical framework for calculating the emission for the different kinds of quantum fields by using the semiclassical approach.  We showed that in the quantum case there exists a nontrivial gravity-matter coupling term of the form $h_{ab}\langle T^{ab}\rangle$. Since all matter fields are coupled to the metric perturbation through their SEM tensor,  all particle species are produced. This was  demonstrated explicitly in the emission formula Eq.~(\ref{1.71}). In contrast, in the classical GR treatment, the radiation from BHs undergoing horizon deformations is emitted only in the from of gravitational waves. We also stress that such a coupling term is absent in the absence of a perturbation, because for  Schwarzschild BHs $T_{ab}=0$ for all matter fields. This discussion highlights the difference between the underlying mechanism that provides the source of energy in the classical case and that in the quantum case. In the classical case, the emission originates from the geometrical properties of the scalar curvature and its associated perturbed metric, which eventually constitute the gravitational SEM tensor of the emitted radiation, $T_{ab}\sim\dot{h}_{ab}^{~2}$. In the quantum case, the source term is the vacuum fluctuation of the matter fields RSEM tensor that is coupled to the perturbed background $h_{ab}$. We also emphasize the importance of the gravity-matter coupling term that appears in Eq.~(\ref{ddd1}) to the resolution of the BHs information paradox.  Whereas in \cite{Giddings:2014nla,Giddings:2017mym} it is shown that operators in this form can transfer information to the outgoing radiation and may eventually lead to unitarization.

			Subsequently, in section (\ref{iqf}), we calculated explicitly the emission of the minimally coupled scalar fields, neutrinos, photons and graviton fields. We find that the flux is always positive, independently on the sign of the deformation parameter $D$, and independent on whether the BH outer surface protrudes outward or inward. Then, by assuming that the emission is approximately that of a black body, we interpreted the additional flux in terms of the BH surface temperature and find the modification to the Hawking temperature. Similarly, the modification was found to be independent of the sign of the horizon displacement parameter $D$.
Later, considering the results of the classical emission we compare the classical and quantum results for the BH luminosity, we conclude that the ratio is controlled by the BH entropy $1/S_{BH}$ and also by the ratio $\Delta t/T_S$, which in some setup could provide significant enhancement factor in Eq.~(\ref{ytr}).

As explained in Section (\ref{DHG}), an external observer can describe the near horizon geometry of a deformed BH in the horizon locking coordinate system. This means  that the horizon position is locked at $r=R_S$, such that $h(R_S)=0$. In this gauge, the perturbed geometry is interpreted in terms of a perturbation in the associated scalar curvature. Alternatively, the change of the Ricci scalar can be described as the deformation of the BH outer horizon with respect to its unperturbed horizon at $r=R_S$, as shown in Fig.~\ref{F}. An external GR observer using the horizon locking coordinate system cannot distinguish classically between a perturbed and an unperturbed BH. However, if she measures the temperature of the unperturbed BH, the result will be $T_H$ and if she measures the temperature of the perturbed BH, then according to Eq.~(\ref{TM}), she finds a different temperature. A possible explanation that the GR observer can provide for the temperature difference is that  there must be additional energy flux originating from the BH interior. Therefore she may deduce that the BH interior is not empty.

		\end{subequations}
		
\section*{Acknowledgments}

The research of RB and YS was supported by the Israel Science Foundation grant no. 1294/16. The research of YS was supported by the BGU Hi-Tech scholarship.. YS thanks the Theoretical Physics Department, CERN for their  hospitality during his visit.

	\appendix
	
\setcounter{equation}{0}
\section{ Gravitational setup}\label{AX}
\renewcommand{\theequation}{\Alph{section}.\arabic{equation}}
In this appendix we explain in detail the connection of tidal deformations  to the discussion that is given in Section~(\ref{BD}). The discussion follows \cite{Poisson:2004cw,Poisson:2009qj,Poisson:2018qqd}. Similar discussions are also presented in \cite{OSullivan:2014ywd,Thorne:1980ru,Alvi:1999cw}.
		
\begin{figure*}[h!]
		\centering
		\includegraphics[scale=.73]{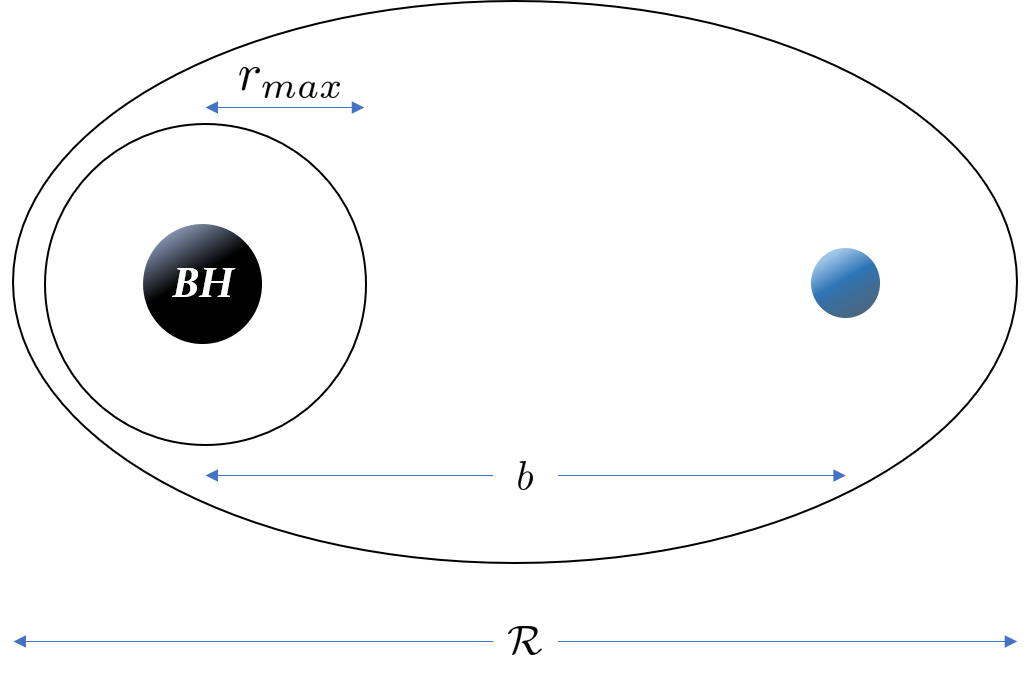}
		\renewcommand{\baselinestretch}{1.0}
		\caption{ { A companion orbiting a BH. The vicinity region of the BH (black sphere) where the BH metric is substantially modified is defined by the black solid circle. The distance $b$ between the companions and radius of curvature $\mathcal{R}>b$ \vspace{-0.2cm} is also shown.  }}
\renewcommand{\baselinestretch}{1.5}\normalsize		
\label{f12}
	\end{figure*}

			To begin, we consider the perturbatons in the horizon-locking gauge about the Schwarzschild backgrounds in the outgoing EF coordinates. The line element is given by
		\begin{gather}
		ds^2~=~-f(r)du^2-2dudr+r^2d\Omega^2~,
		\label{ED}
		\end{gather}	
		were $f(r)=1-2M/r$. The metric perturbation $h_{\mu\nu}$ is given as an expansion in $r/\mathcal{R}$ where $\mathcal{R}$ is the radius of curvature. The scale $\mathcal{R}$ defines the region in space where the BH gravitational field interacts with the tidal field of an external object. As explained in Section (\ref{BD}), to guarantee a weak gravitational interaction, the Schwarzschild radius $R_S$ of the BH must satisfy $R_S\ll\mathcal{R}$. For the case of a binary system, where the external object has mass $M'$ and the separation distance is $b$, the radius of curvature is given by $\mathcal{R}^2\sim b^3/(M+M')$. This is illustrated in Fig.~\ref{f12}.
		Here, we list the leading terms which is second order in $r/\mathcal{R}$.  Then, by imposing the proper gauge conditions, in the vicinity of the BH the metric perturbation takes the form
		\begin{align}
		&h_{uu}~=~-r^2f(r)^2{C}_{2}+O(r^3/\mathcal{R}^3)\label{HUU}~,\\
		&h_{ur}~=~0\label{HUR}~,\\
		&h_{uA}~=~\dfrac{2}{3}r^3f(r)\left({C}_{2}^A+{B}_{2}^A\right)+O(r^4/\mathcal{R}^3)\label{HUA}~,\\
		&h_{AB}~=~-\dfrac{1}{3}r^4\left[\left(1-\dfrac{2M^2}{r^2}\right){C}_{2}^{AB}+{B}_{2}^{AB}\right]+O(r^3/\mathcal{R}^4)\label{HAB}~,
		\end{align}
		where the ${C}_{2}, {C}^A_{2},{B}^A_{2}, {C}_{2}^{AB},{B}_{2}^{AB}$ are the dimensionful tidal quadrupole moments  of the scalar, vector and tensor spherical harmonics,   respectively. We later define the appropriate scalar harmonics. The definition of the vector and tensor harmonics is given in \cite{Poisson:2004cw} and is not relevant to our purpose, since, as previously mentioned, they do not contribute to the emitted flux. The perturbation strength scales as $	{C}\sim{B}\sim \mathcal{R}^{-2}$. The contribution of the higher-order terms: the octupole moments that scale as $r^3/\mathcal{R}^3$ and the hexadecapole moments that scale as $r^4/\mathcal{R}^4$ are smaller than the quadrupole moments that scales as $r^2/\mathcal{R}^2$. We therefore consider only the contribution of the quadrupole term $l=2$. In the limit $r/\mathcal{R}\rightarrow 0$, while keeping $M/r$ fixed, the metric perturbation
		vanishes.
		
		One could also define the dimensionless parameters
		\begin{gather}
		\widetilde{C}\sim\widetilde{B}\sim \dfrac{r^2}{\mathcal{R}^2}~,
		\label{EP}\\
		\widetilde{D}\sim \dfrac{M^2}{\mathcal{R}^2}~,
		\label{MP}
		\end{gather}
		where the $\widetilde{C},\widetilde{B}$ denote the expansion parameter of the metric perturbation. The parameter $\widetilde{D}$ can be viewed as the expansion parameter in the vicinity of the horizon $r\sim M$. It is related to the tidal fields by $\widetilde{D}= \widetilde{C}M^2/r^2 $ and, as we show below, it is useful for the  description of the horizon deformation.
		The scalar harmonics of the tidal fields are defined by
		\begin{gather}
		{C}_{2}~=~C \sum_mY_{2m}. \label{SHT}\\
		\widetilde{D}_{2}~=~M^2{C}\sum_mY_{2m}, \label{SHT1}
		\end{gather}
		where $C\sim \mathcal{R}^{-2}$ and its exact value is determined by defining the components of the Weyl tensor \cite{Poisson:2004cw}. For the purpose of this work it is unnecessary to specify them further, since our interest is in the region $2M<r<r_{max}$, where  $r_{max}\ll\mathcal{R}$ Fig.~\ref{f12}. Then, the construction of the Weyl tensor and its associated tidal fields will only indicates how small  the ratio $\widetilde{C}\sim r^2/\mathcal{R}^2\ll1$ is. The important point is that this region is finite and the perturbative description in Eqs.~(\ref{HUU})-(\ref{HAB}) is valid only for $r<r_{max}$. In addition, it is shown in \cite{Taylor:2008xy,Alvi:1999cw} that in regions where $r_{max}<r<\mathcal{R}$ the metric takes the form of post-Newtonian expansion, such that the induced corrections to the BH spacetime are significantly smaller than those that are given in Eqs.~(\ref{HUU})-(\ref{HAB}). Therefore, one is mainly\vspace{0.2cm} interested in the region $R_S<r<r_{max}$.
		
		The spherical harmonics listed above are defined as the real part of the $Y_{2m}$,
		\renewcommand{\baselinestretch}{1.1}\normalsize
		\begin{equation}
		\begin{split}
		& Y_{2,0~}~=~1-3cos^2\theta~,\\
		& Y_{2,1c}~=~2cos\theta sin\theta cos\phi~, \\
		& Y_{2,1s}~=~2cos\theta sin\theta sin\phi~,  ~~~~~~~~~~~~~~~~~~~~~m=0, 1c, 1s, 2c, 2s\\
		& Y_{2,2c}~=~sin^2\theta cos2\phi~,\\
		&Y_{2,2s}~=~sin^2\theta sin2\phi~.\label{SPT}
		\end{split}
		\end{equation}
		\renewcommand{\baselinestretch}{1.5}\normalsize
		We also define the characteristic time scale $\mathcal{T}\sim \sqrt{b^3/M}\sim \mathcal{R}$ and the frequency $\omega\sim\mathcal{T}^{-1} \sim \widetilde{D}^{1/2}/M$, where $\mathcal{T}\gg R_s$ describes a slowly changing quadrupole. For example, in a binary system with externally moving object with mass $M'$ the frequency is given by $\omega^2=(M+M')/b$. The time derivative of the quadrupole moments, is given by
		\begin{gather}
		\dot{\widetilde{C}}\sim{\widetilde{C}}{\mathcal{T}}^{-1}\sim \widetilde{C}\omega \sim{\widetilde{C}}\dfrac{\widetilde{D}^{1/2}}{M} \sim\dfrac{r^2}{\mathcal{R}^3}~.
		\label{TS}
		\end{gather}
		
		Another important scale is the inhomogeneity scale $\mathcal{L}$, which measures the degree of spatial variation of the induced tidal fields. In the context of a binary system, the analogue scale is the relative distance between the constituents $\mathcal{L}\sim b$. It is the smallest scale among $\mathcal{T}, \mathcal{R}$, so $\mathcal{L} < \mathcal{R}$ and given by
		$	\mathcal{L}\sim {M}{\widetilde{D}^{-1/3}}$. The definition of $r_{max}$ is given in terms of the length scale $\mathcal{L}$ and the cutoff $\mathcal{R}$. First, it is clear that $r_{max}$ has to be smaller than both $\mathcal{L}$ and $\mathcal{R}$, this indicates that it must be determined according to the smallest scale in the problem, then $r_{max}\lesssim \alpha \mathcal{L}$ with $\alpha$ being a dimensionless parameter $\alpha<1$. Then, the only way to construct $\alpha$ from $\mathcal{L,\mathcal{R}}$ is using the relation $\alpha\sim \mathcal{L}/\mathcal{R}$. Thus $r_{max}\sim\mathcal{L}^2/\mathcal{R}\sim\sqrt{M \mathcal{L}}\sim M\widetilde{D}^{-1/6}$, which agrees with \cite{Taylor:2008xy}.
		For completeness, we stress that the metric perturbation in Eqs.~(\ref{HUU})-(\ref{HAB}) is well defined in the region $R_s < r  \lesssim M{\widetilde{D}}^{-1/6}$. Otherwise, at larger distances, the expansion parameter is too large and the perturbative treatment breaks down.

		In accordance with the above relations, we now wish to express the horizon shift $\Delta R$ as the result of the tidal deformation described by the metric perturbation. As explained in Section~(\ref{DHG}) the explicit expression of the Ricci scalar \cite{Vega:2011ue}, is given by
		\begin{gather}
			\mathcal{R}~=~\dfrac{1}{2M^2}\left(1-\dfrac{1}{6}l(l-1)(l+1)(l+2)\widetilde{D}_{lm}\right)~.
			\label{RS}
		\end{gather}
		The BH horizon radial deviation, following \cite{Vega:2011ue,OSullivan:2014ywd} is given by
		\begin{gather}
		R_{lm}~=~R_{S}\left(1-\dfrac{1}{6}l(l+1)\widetilde{D}_{lm}\right)~.
		\end{gather}
		The radial shift $\Delta R_{2}=R_{2}-R_S$ then reads
		\begin{gather}
		\Delta R_{2}~=~-\widetilde{D}_{2}R_S~.\label{DR}
		\end{gather}
		So the BH horizon extends from its original unperturbed location up to a distance scale that is determined by the factor $\Delta R_s/R_s = -\widetilde{D}$.
		
		We can now identify $\widetilde{D}$ with the horizon displacement parameter $D$ in Eq.~(\ref{Rlm}), $	\widetilde{D}_{2}=-D_{2}$, which establishes the connection between the description in Section~(\ref{DHG}) and the original setup of the tidal deformation.

\end{document}